\documentstyle[12pt, a4]{article}

\begin{document}
\author{Ramazan Ko\c{c}, Okan \"{O}zer\footnote{Corresponding Author: ozer@gantep.edu.tr} and Hayriye
T\"{u}t\"{u}nc\"{u}ler \and Department of Engineering Physics,
University of Gaziantep, \and 27310 Gaziantep-T\"{u}rkiye}
\title{Solution of the Matrix Hamiltonians via asymptotic iteration method}
\maketitle
\begin{abstract}
A method is suggested to obtain solutions of the various quantum
optical Hamiltonians in the framework of the asymptotic iteration
method. We extend the notion of asymptotic iteration method to
solve the $2\times 2$ matrix Hamiltonians. On a particular case,
eigenvalues of the Rabi and Rashba Hamiltonians are computed. The
method presented here reproduces a number of earlier results in a
natural way as well as leads to a novel findings. Possible
generalizations of the method are also suggested.
\end{abstract}

{\bf Pacs Numbers}: 03.65.Ge, 03.65.Ca, 73.21.La


\newpage\

\section{Introduction}

The solutions of the quantum optical Hamiltonians are an important theme in
the existing literature \cite{prinz,wolf,winkler,wang,band}. There has been
a great deal of interest in quantum optical models which reveal new physical
phenomena described by the $2\times 2$ matrix Hamiltonians. These
Hamiltonians appear at the different fields of the physics \cite%
{rashba,das,tutunculer,judd,dresselhaus,jaynes}. It is well known
that the interaction of a two-level systems with a radiation field
has a long varied history and these problems are often modelled by
using $2\times 2$ matrix Hamiltonians. As an example a physical
system which describes the optical and electrical properties of
confined electrons in semiconductor quantum wells, quantum dots
and quantum wires depends on the Rashba spin orbit coupling and
its equation is given by $2\times 2$ matrix Hamiltonian. The
Rashba splitting has been observed in many experiments and it
constitutes the basis of the proposed electronic nano structures
\cite{rashbaei}. Due to the practical and technological importance
of these models, it is not surprising that various aspects have
been studied both analytically and numerically. Such systems have
often been analyzed by using numerical methods because the
implementation of the analytical techniques does not yield simple
analytical expressions. Remarkably, exact solutions have not been
thus far presented except for special cases even though it has
been suggested that the problem may be solved exactly and their
analytical treatments require tedious calculations
\cite{tur,tsit,frus,koc1,koc2,gover,koc3,reik,loorits}.

In recent years much attention has been focused on asymptotic
iteration method (AIM)
\cite{ciftci1,sous,ciftci2,ciftci3,barakat1,amore,bayrak}. This
method reproduces exact solutions to many exactly solvable
differential equations and these equations can be related to the
Schr\"{o}dinger equation. It also gives accurate results for the
non solvable Schr\"{o}dinger equation including sextic oscillator,
cubic oscillator, deformed Coulomb potential etc., which are
important in applications to many problems in physics. Encouraged
by its satisfactory performance through comparisons with the other
methods, we feel tempted to develop AIM to solve matrix
differential equations. In contrast to the solution of the
Schr\"{o}dinger equation by using AIM including Coulomb, Morse,
harmonic oscillator, etc. type potentials, the study of the
quantum optical Hamiltonians has not attracted much attention in
the literature. Such Hamiltonians have been found to be useful in
the study of electronic properties of semiconductors, quantum dots
and quantum wells.

The aim of this paper is to develop AIM for solving $2\times \ 2$ matrix
Hamiltonians and discuss their applications. Results of our procedure
include the solutions of the Rabi \cite{jaynes} and Rashba \cite{rashba}
Hamiltonians. We provide a first step toward the extension of the technique
to the solution of the various matrix Hamiltonians, whose spectrum can not
be obtained exactly.


The paper is organized as follows. In section 2, we develop the
AIM to obtain eigenvalues and eigenfunctions of the wide range of
the matrix Hamiltonians. Section 3 is devoted to solve the Rabi
Hamiltonian in the framework of the AIM. In section 4 we present
the solution of the Rashba Hamiltonian. In section 3 and 4 we also
discuss the bosonisation of the physical Hamiltonians whose
original forms are given as differential operators. We also
present a procedure to transform the bosons in to Bargman-Fock
space which is necessary to obtain first-order matrix differential
Hamiltonians. The paper ends with a brief conclusion.

\section{Formalism of the asymptotic iteration method for matrix Hamiltonians}

The AIM is proposed to solve the second-order differential
equations and the details can be found in \cite{ciftci1}. In this
section we systematically extend the method for the $2\times \ 2$
first-order matrix differential equations. We begin by rewriting a
first-order differential equation in the following matrix form:
\begin{equation}
I\phi ^{\prime }=u_{0}\phi  \label{umatrix}
\end{equation}%
where $\phi =\left[ \phi _{1},\phi _{2}\right] ^{T}$,
two-component column vector $u_{0}$ is $2\times \ 2$ matrix
function and $I$ is $2\times \ 2$ unit matrix. Note that $\phi$ is
the function of $x$ and $\phi ^{\prime }$ is the first derivative
with respect to $x$. Now, in order to obtain a general solution to
this equation in the framework of the AIM we use the similar
arguments given in \cite{ciftci1}. More explicitly, the
differential equation (\ref{umatrix}) can be written as the two
coupled equations
\begin{equation}
\phi _{1}^{\prime }=a_{0}\phi _{1}+b_{0}\phi _{2};\quad \phi
_{2}^{\prime }=c_{0}\phi _{2}+d_{0}\phi _{1}  \label{coupledeq}
\end{equation}%
where $a_{0},b_{0},c_{0}$ and $d_{0}$ are elements of the matrix $u_{0}$. It
is easy to show that $n^{th}$ derivative of the $\phi _{1}$ and $\phi _{2}$
can be expressed as
\begin{eqnarray}
\phi _{1}^{\prime \prime } &=&a_{1}\phi _{1}+b_{1}\phi _{2};\quad \phi
_{2}^{\prime \prime }=c_{1}\phi _{2}+d_{1}\phi _{1}  \nonumber \\
\phi _{1}^{\prime \prime \prime } &=&a_{2}\phi _{1}+b_{2}\phi _{2};\quad
\phi _{2}^{\prime \prime \prime }=c_{2}\phi _{2}+d_{2}\phi _{1}  \nonumber \\
&&\cdots  \label{nthderivation} \\
\phi _{1}^{(n)} &=&a_{n-1}\phi _{1}+b_{n-1}\phi _{2};\quad \phi
_{2}^{(n)}=c_{n-1}\phi _{2}+d_{n-1}\phi _{1}  \nonumber \\
\phi _{1}^{(n+1)} &=&a_{n}\phi _{1}+b_{n}\phi _{2};\quad \phi
_{2}^{(n+1)}=c_{n}\phi _{2}+d_{n}\phi _{1}.  \nonumber
\end{eqnarray}%
In order to discuss the asymptotic properties of (\ref{umatrix}),
it is necessary to determine the coefficients $a_{n},b_{n},c_{n}$
and $d_{n}.$ After some straightforward calculation one can obtain
the following relations:
\begin{eqnarray}
a_{n} &=&a_{0}a_{n-1}+a_{n-1}^{\prime }+d_{0}b_{n-1}  \nonumber \\
b_{n} &=&b_{0}a_{n-1}+b_{n-1}^{\prime }+c_{0}b_{n-1}  \nonumber \\
c_{n} &=&c_{0}c_{n-1}+c_{n-1}^{\prime }+b_{0}d_{n-1}  \label{constr1} \\
d_{n} &=&d_{0}c_{n-1}+d_{n-1}^{\prime }+a_{0}d_{n-1}.  \nonumber
\end{eqnarray}%
Our task is now to introduce the asymptotic aspect of the method. For this
purpose $n^{th}$ and $(n+1)^{th}$ derivative of the $\phi _{1}$ and $\phi
_{2}$ can be written as%
\begin{eqnarray}
\phi _{1}^{(n)} &=&a_{n-1}\left( \phi _{1}+\frac{b_{n-1}}{a_{n-1}}\phi
_{2}\right) ;\quad \phi _{2}^{(n)}=c_{n-1}\left( \phi _{2}+\frac{d_{n-1}}{%
c_{n-1}}\phi _{1}\right)  \nonumber \\
\phi _{1}^{(n+1)} &=&a_{n}\left( \phi _{1}+\frac{b_{n}}{a_{n}}\phi
_{2}\right) ;\quad \phi _{2}^{(n+1)}=c_{n}\left( \phi _{2}+\frac{d_{n}}{c_{n}%
}\phi _{1}\right) .  \label{nplus1deriv}
\end{eqnarray}%
The coefficients $a_{0},b_{0},$ $d_{0}$ and $c_{0}$ include the coupling
constants. Therefore, for sufficiently large $n$ we can suggest the
following asymptotic constraints:%
\begin{equation}
\frac{b_{n-1}}{a_{n-1}}=\frac{b_{n}}{a_{n}}=\gamma _{1}; \quad
\frac{d_{n-1}}{c_{n-1}}=\frac{d_{n}}{c_{n}}=\gamma _{2}.
\label{qcond}
\end{equation}%
In this formalism the relations given in (\ref{qcond}) imply that
the wave function $\phi _{1}$ and $\phi _{2}$ are truncated for
sufficiently large $n$ and the roots of the relations
(\ref{qcond}) belong to the spectrum of the matrix Hamiltonian.
Therefore one can easily compute the eigenenergies of the
Hamiltonian by solving (\ref{qcond}) for the energy term when
$x\rightarrow x_{0}$.

Under the asymptotic condition given in (\ref{qcond}), one can
find the wave functions $\phi _{1}$ and $\phi _{2}$. When we take
$\frac{\phi _{1}^{(n+1)}}{\phi _{1}^{(n)}}$ and
$\frac{\phi_{2}^{(n+1)}}{\phi_{2}^{(n)}}$ by using
(\ref{nplus1deriv}) under the constraints given in (\ref{qcond}),
we obtain:
\begin{equation}
\phi _{1}^{(n)}=\exp \left( \int \frac{a_{n}}{a_{n-1}}dx\right)
\quad or \quad \phi _{2}^{(n)}=\exp \left( \int
\frac{c_{n}}{c_{n-1}}dx\right) . \label{phi1phi2}
\end{equation}
Substituting the expression of $a_{n}$ (and $c_{n}$) given in
(\ref{constr1}) into (\ref{phi1phi2}) and then replacing the
$\phi_{1}^{(n)}$ (and $\phi_{2}^{(n)} $) in the first expression
of (\ref{nplus1deriv}), respectively, one gets the following
expressions;
\begin{equation}
\phi _{1}+\gamma_{1}\phi _{2}=\exp \left( \int \left(
a_{0}+\gamma_{1}d_{0}\right) dx\right)\quad or \quad
\phi_{2}+\gamma_{2}\phi _{1}=\exp \left( \int \left(
c_{0}+\gamma_{2}b_{0}\right) dx\right) .  \label{wavefunctions}
\end{equation}

An immediate practical consequence of these results is that the
eigenvalues and eigenfunctions of the various quantum optical
Hamiltonians can easily be determined. We mention here that there
exist numerous physical Hamiltonians which can be written in the
form of the first-order matrix differential equation or can be
transformed in the form of the first-order matrix differential
equation. For example, the Dirac equation \cite{ciftci3} is a
first-order matrix differential equation while most of the
Hamiltonians of the quantum optical systems can be written as
first-order matrix differential equation in the Bargman-Fock
space. Now we can determine the eigenvalues and eigenfunctions of
the $2\times 2$ matrix Hamiltonians. In the following sections, it
will be shown that this asymptotic approach opens the way to
treatment of a large class of matrix Hamiltonians of practical
interest.

\section{Solution of the Rabi Hamiltonian by using AIM}

In this section we take a new look at the solution of Rabi
Hamiltonian through the AIM. The Rabi Hamiltonian is a successful
model of the interaction between matter and electromagnetic
radiation. The Hamiltonian describes interacting a dipole
interaction with a single mode of radiation in a two-level system.
This Hamiltonian can be solved exactly within the framework of
rotating wave approximation and its quasi-exact or isolated exact
solutions can be obtained for some specific cases. Let us consider
the Rabi Hamiltonian for a two-level atom \cite{jaynes}:
\begin{equation}
H=-\frac{\hbar ^{2}}{2m}\frac{d^{2}}{dx^{2}}I+\frac{1}{2}m\omega
^{2}x^{2}I+\omega _{0}\sigma _{0}+\kappa \left( \sigma _{+}+\sigma
_{-}\right) x  \label{RabiH}
\end{equation}
where $\omega _{0}$ is the atomic level splitting, $\omega $ is the
frequency of the oscillation and $\kappa $ is the coupling strength of the
atom to the field and, $\sigma _{0},\sigma _{+},\sigma _{-}$ are
\begin{equation}
\sigma _{0}=\left(
\begin{array}{cc}
-1 & 0 \\
0 & 1%
\end{array}%
\right) ;\sigma _{+}=\left(
\begin{array}{cc}
0 & 1 \\
0 & 0%
\end{array}%
\right) ;\sigma _{-}=\left(
\begin{array}{cc}
0 & 0 \\
1 & 0%
\end{array}%
\right)  \label{paulimatrices}
\end{equation}%
Pauli-related matrices. Eigenvalues of the Hamiltonian
(\ref{qcond}) can be obtained from the eigenvalue equation
$H\phi=E\phi$ where $\phi $\ is a two component wave function. In
general, it is difficult to determine the asymptotic aspects of
the (\ref{RabiH}) in the present form. Therefore it is worth to
transform (\ref{RabiH}) to an appropriate form. One way to
transform a Hamiltonian in the form of the first-order matrix
differential equation is to construct its bosonic representation.
We are interested in the two-level system in one and
two-dimensional geometry, whose Hamiltonians are given in terms of
bosons-fermions or matrix-differential equations. Therefore, it is
worth to express a suitable differential realizations of the
bosons. Consider the following boson realizations
\begin{equation}
a=\sqrt{\frac{\hbar }{2m\omega }}\frac{d}{dx}+\sqrt{\frac{m\omega
}{2\hbar }}x;\quad
a^{+}=-\sqrt{\frac{\hbar }{2m\omega
}}\frac{d}{dx}+\sqrt{\frac{m\omega }{2\hbar }}x.  \label{aadiger}
\end{equation}
They satisfy the usual commutation relation
\begin{equation}
\left[ a,a^{+}\right] =1;\quad \left[ a,a\right] =\left[
a^{+},a^{+}\right] =0  \label{commrel}
\end{equation}
In terms of the boson operators, the Hamiltonian (\ref{RabiH}) can
be written as
\begin{equation}
H=\hbar \omega \left( a^{+}a+\frac{1}{2}\right) I+\omega
_{0}\sigma _{0}+ \sqrt{\frac{\hbar }{2m\omega }}\kappa \left(
\sigma _{+}+\sigma _{-}\right) (a^{+}+a).  \label{BosonicH}
\end{equation}
It is interesting that this type Hamiltonian gives good results
when it is solved by using the AIM in the Bargman-Fock space. In
order to transform the Hamiltonian to the Bargman-Fock space, we
introduce the following transformation operator:
\begin{equation}
\Gamma _{1}=\exp \left[ \frac{\alpha _{1}}{2}\left(
a^{2}+a^{+2}\right)\right] .  \label{Gamma1}
\end{equation}
The action of the operator on the bosons is given by
\begin{equation}
\Gamma _{1}a\Gamma _{1}^{-1}=a\cos \alpha _{1}-a^{+}\sin \alpha
_{1};\Gamma _{1}a^{+}\Gamma _{1}^{-1}=a^{+}\cos \alpha _{1}+a\sin
\alpha _{1}.\quad \label{GaG}
\end{equation}
For the values $\alpha _{1}=\pi /4$ boson operators take the form
\begin{equation}
a\rightarrow \frac{1}{\sqrt{2}}\left( a-a^{+}\right)
=\sqrt{\frac{\hbar }{m\omega }}\frac{d}{dx};\quad a^{+}\rightarrow
\frac{1}{\sqrt{2}}\left( a+a^{+}\right) =\sqrt{\frac{m\omega
}{\hbar }}x  \label{agoesto}
\end{equation}
The final form of the Hamiltonian (\ref{BosonicH}) with the
realization (\ref{agoesto}) can be expressed as
\begin{equation}
H=\hbar \omega \left( x\frac{d}{dx}+\frac{1}{2}\right) I+\omega
_{0}\sigma _{0}+\frac{\kappa }{\sqrt{2}}\left( \sigma _{+}+\sigma
_{-}\right) \left( x+ \frac{\hbar }{m\omega }\frac{d}{dx}\right) .
\label{finalH}
\end{equation}
It is obvious that the Hamiltonian (\ref{finalH}) can easily be
written as the two coupled first-order differential equation
\begin{eqnarray}
\phi _{1}^{\prime } &=&a_{0}\phi _{1}+b_{0}\phi _{2}  \nonumber \\
\phi _{2}^{\prime } &=&c_{0}\phi _{2}+d_{0}\phi _{1}.
\label{phiderivs}
\end{eqnarray}%
For the sake of simplicity let us take $\hbar =\omega =m=1$ at
that point, then the coefficients of the coupled differential
equations are given by
\begin{eqnarray}
a_{0} &=&\frac{x\left( 2E-1+\kappa ^{2}-2\omega _{0}\right) }{2x^{2}-\kappa
^{2}}  \nonumber \\
b_{0} &=&\frac{\kappa \left( 1-2E-2x^{2}-2\omega _{0}\right) }{\sqrt{2}%
\left( 2x^{2}-\kappa ^{2}\right) }  \nonumber \\
c_{0} &=&\frac{x\left( 2E-1+\kappa ^{2}+2\omega _{0}\right) }{2x^{2}-\kappa
^{2}}  \label{coeffs1} \\
d_{0} &=&\frac{\kappa \left( 1-2E-2x^{2}+2\omega _{0}\right) }{\sqrt{2}%
\left( 2x^{2}-\kappa ^{2}\right) }.  \nonumber
\end{eqnarray}

Using a simple MATHEMATICA program one can compute
$a_{n},b_{n},c_{n}$ and $d_{n}$ by the relations given in
(\ref{constr1}). On the other hand, for each iteration, the
quantization condition
$\delta_{1}(x)=b_{n-1}(x)a_{n}(x)-a_{n-1}(x)b_{n}(x)$ (and
$\delta_{2}(x)=d_{n-1}(x)c_{n}(x)-c_{n-1}(x)d_{n}(x)$) depends on
some variables namely $E_{n},\kappa ,\omega _{0}$ and $x$. It is
noticed that the iterations should be terminated by imposing the
condition $\delta_{i}(x)=0, i=1,2$ as an approximation to Eq.
(\ref{qcond}) to obtain the eigenenergies. The calculated
eigenenergies $E_{n}$ by means of this condition should, however,
be independent of the choice of $x$. The choice of $x$ is observed
to be critical only to the speed of the convergence of the
eigenenergies, as well as for the stability of the process. In our
study it has been observed that the optimal choice for $x$ is the
extremum point of the potential that is when $x=0$. Therefore, we
set $x=0$ at the end of the iterations. Then, the roots of the
iteration produce eigenenergies for the Rabi Hamiltonian. The
results are reported in Table \ref{tab:a}.


We can also obtain eigenfunctions of the Hamiltonian (\ref{RabiH})
by using the relation (\ref{wavefunctions}). \ Analytical
expression for the ground state wave function of the Rabi
Hamiltonian for some different parameters are given by
\begin{eqnarray}
\kappa &=&0;\quad \omega _{0}=0;\quad E=n+\frac{1}{2};\quad \phi
=\left(
\begin{array}{c}
C_{1}x^{n} \\
C_{2}x^{n}
\end{array}
\right) ;
\nonumber \\
\kappa &=&0;\quad \omega _{0}=\frac{1}{2};\quad E=n;\quad \phi
=\left(
\begin{array}{c}
C_{1}x^{n-1} \\
C_{2}x^{n}
\end{array}
\right) ;  \label{kappas}  \\
\kappa &=&\frac{1}{2};\quad \omega _{0}=0;\quad
E=n+\frac{3}{8};\quad \phi =\left(
\begin{array}{c}
C_{1}e^{-\frac{x}{2\sqrt{2}}}(1+2\sqrt{2}x)^{n} \\
C_{1}e^{-\frac{x}{2\sqrt{2}}}(1+2\sqrt{2}x)^{n}
\end{array}
\right).\nonumber
\end{eqnarray}
The solution of this system describes a quantum mechanical state
of $H$ provided that $\phi $ belongs to the Bargman-Fock space. It
is noticed that when $\kappa $ or $\omega _{0}$ is zero then the
Hamiltonian (\ref{RabiH}) can be solved exactly. We have observed
that for $n=5$, accurate eigenvalues can be obtained after 15
iteration. As a consequence we have demonstrated that the solution
of the Rabi Hamiltonian can be treated within the AIM. Our
approach is relatively simple and gives accurate result.

\section{Solution of the Rashba Hamiltonian by using AIM}

The origin of the Rashba spin-orbit coupling in quantum dots is due to the
lack of inversion symmetry which causes a local electric field perpendicular
to the plane of heterostructure. The Hamiltonian representing the Rashba
spin orbit coupling for an electron in a quantum dot can be expressed as
\cite{rashba}%
\begin{equation}
H_{R}=\frac{\lambda _{R}}{\hbar }\left( p_{y}\sigma
_{x}-p_{x}\sigma _{y}\right)  \label{RashbaH}
\end{equation}
where $\lambda _{R}$ represents the strength of the spin orbit
coupling, and it can be adjusted by changing the asymmetry of the
quantum well via external electric field and the matrices $\sigma
_{x},$ and $\sigma _{y}$ are Pauli matrices. Now we assume that
the electron is confined in a
parabolic potential%
\begin{equation}
V=\frac{1}{2}m^{\ast }\omega _{0}^{2}(x^{2}+y^{2})
\label{parabolicpot}
\end{equation}
here $m^{\ast}$ is the effective mass of the electron and
$\omega_{0}$ is the confining potential frequency. The Hamiltonian
describing an electron in two-dimensional quantum dot takes the
form
\begin{equation}
H=\frac{1}{2m^{\ast }}\left( P_{x}^{2}+P_{y}^{2}\right)
+\frac{1}{2}g\mu B\sigma _{0}+V+H_{R}.\label{twodimHam}
\end{equation}
The term $\frac{1}{2}g\mu B\sigma _{0}$ introduces the Zeeman
splitting between the $(+)x-$polarized spin up and
$(-)x-$polarized spin down. The factors $g$ is the gyromagnetic
ratio and $\mu $ is the Bohr magneton. The kinetic momentum ${\bf
P=p+eA}$ is expressed with canonical momentum ${\bf p}=-i\hbar
(\partial _{x},\partial _{y},0)$ and the vector potential ${\bf
A}$ can be related with the magnetic field ${\bf B=\nabla }\times
{\bf A}$. The choice of symmetric gauge vector potential ${\bf
A}=B/2(-y,x,0)$, leads to the following Hamiltonian
\begin{eqnarray}
H &=&-\frac{\hbar ^{2}}{2m^{\ast }}\left( \frac{\partial
^{2}}{\partial x^{2}}+\frac{\partial ^{2}}{\partial y^{2}}\right)
+\frac{1}{2}m^{\ast }\omega
^{2}(x^{2}+y^{2})  \nonumber \\
&&+\frac{1}{2}i\hbar \omega _{c}\left( y\frac{\partial }{\partial
x}-x\frac{\partial }{\partial y}\right) +\frac{1}{2}g\mu B\sigma
_{0}+H_{R} \label{gaugeham}
\end{eqnarray}
where $\omega_{c}=eB/m^{\ast}$ stands for the cyclotron frequency
of the electron, $\omega =\sqrt{\omega _{0}^{2}+\left(\frac{\omega
_{c}}{2}\right)^{2}}$ is the effective frequency. From now on we
restrict ourselves to the solution of (\ref{gaugeham}).

\subsection{Bosonisation of the Hamiltonian}

The Hamiltonian (\ref{gaugeham}) can not be solved within the
framework of the AIM in the present form. Our task is now to
demonstrate that the Hamiltonian (\ref{gaugeham}) can be expressed
as two coupled first-order differential equation in the
Bargman-Fock space. One way to express the Hamiltonian $H$ with
boson operators is to introduce an appropriate differential
realization for bosons. It can easily be bosonised when the boson
operators are realized
as%
\begin{eqnarray}
a^{+} &=&\sqrt{\frac{m^{\ast }\omega }{4\hbar }}(x+iy)-\sqrt{\frac{\hbar }{%
4m^{\ast }\omega }}(\partial _{x}+i\partial _{y})  \nonumber \\
a &=&\sqrt{\frac{m^{\ast }\omega }{4\hbar }}(x-iy)+\sqrt{\frac{\hbar }{%
4m^{\ast }\omega }}(\partial _{x}-i\partial _{y})  \nonumber \\
b^{+} &=&\sqrt{\frac{m^{\ast }\omega }{4\hbar }}(x-iy)-\sqrt{\frac{\hbar }{%
4m^{\ast }\omega }}(\partial _{x}-i\partial _{y})  \label{bosonoperators} \\
b &=&\sqrt{\frac{m^{\ast }\omega }{4\hbar }}(x+iy)+\sqrt{\frac{\hbar }{%
4m^{\ast }\omega }}(\partial _{x}+i\partial _{y}).  \nonumber
\end{eqnarray}%
They satisfy the usual commutation relations. Insertion of
(\ref{bosonoperators}) into (\ref{gaugeham}) yields the following
Hamiltonian:
\begin{eqnarray}
H &=&\hbar \omega (a^{+}a+b^{+}b+1)+\frac{\hbar \omega _{c}}{2}\left(
a^{+}a-b^{+}b\right)  \nonumber \\
&&-\sqrt{\frac{m^{\ast }\omega }{\hbar }}\lambda _{R}\left[
(b^{+}-a)\sigma _{+}+(b-a^{+})\sigma _{-}\right] +\frac{1}{2}g\mu
B\sigma _{0}  \label{newham}
\end{eqnarray}%
Now we turn our attention to the transformation of the Hamiltonian
in the form of the first-order one variable matrix differential
equation. A simple connection between the Hilbert space and the
Bargman-Fock space can be obtained by transforming the
differential realizations of the creation and annihilation
operators (\ref{bosonoperators}). This can be done by introducing
the following similarity transformation operators
\begin{equation}
\Gamma _{2}=\exp \left[ \frac{\alpha _{2}}{2}\left(
b^{2}+b^{+2}\right)\right] ;\quad \Lambda =\exp \left[ \beta
\left( a^{+}b+b^{+}a\right) \right] .  \label{Gamma2}
\end{equation}%
The operators act on the bosons as follows:%
\begin{eqnarray}
\Gamma _{1}a\Gamma _{1}^{-1} &=&a\cos \alpha _{1}-a^{+}\sin \alpha
_{1};\quad \Gamma _{1}a^{+}\Gamma _{1}^{-1}=a^{+}\cos \alpha _{1}+a\sin
\alpha _{1}  \nonumber \\
\Gamma _{2}b\Gamma _{2}^{-1} &=&a\cos \alpha _{2}-a^{+}\sin \alpha
_{2};\quad \Gamma _{2}a^{+}\Gamma _{2}^{-1}=a^{+}\cos \alpha _{2}+a\sin
\alpha _{2}\quad  \nonumber \\
\Lambda a\Lambda ^{-1} &=&a\cos \beta -b\sin \beta ;\quad \Lambda
a^{+}\Lambda ^{-1}=a^{+}\cos \beta -b^{+}\sin \beta  \label{actsonbosons} \\
\Lambda b\Lambda ^{-1} &=&b\cos \beta +a\sin \beta ;\quad \Lambda
b^{+}\Lambda ^{-1}=b^{+}\cos \beta +a^{+}\sin \beta  \nonumber
\end{eqnarray}%
These transformations play a key role to construct one variable
first-order matrix differential equation form of
(\ref{newham}).The similar transformation and the change of the
variable $y\rightarrow iy$ gives the
following realizations%
\begin{eqnarray}
a &\rightarrow &\Lambda \Gamma _{1}a\Gamma _{1}^{-1}\Lambda ^{-1}=\sqrt{%
\frac{\hbar }{m^{\ast }\omega }}\frac{\partial }{\partial x},\quad
a^{+}\rightarrow \Lambda \Gamma _{1}a^{+}\Gamma _{1}^{-1}\Lambda ^{-1}=\sqrt{%
\frac{m^{\ast }\omega }{\hbar }}x  \nonumber \\
b &\rightarrow &\Lambda \Gamma _{2}b\Gamma _{2}^{-1}\Lambda ^{-1}=\sqrt{%
\frac{\hbar }{m^{\ast }\omega }}\frac{\partial }{\partial y},\quad
b^{+}\rightarrow \Lambda \Gamma _{2}b^{+}\Gamma _{2}^{-1}\Lambda ^{-1}=\sqrt{%
\frac{m^{\ast }\omega }{\hbar }}y.  \label{abgoesto}
\end{eqnarray}%
It is obvious that the Hamiltonian (\ref{newham}) can be put in
the form of the first-order matrix differential with the
realization of bosons (\ref{abgoesto}), but it consists of two
variables $x$ and $y$. In order to separate the variables consider
the following conserved quantity of the Hamiltonian
(\ref{newham}):
\begin{equation}
K=a^{+}a-b^{+}b-\frac{1}{2}\sigma _{0}.  \label{Kspace}
\end{equation}
If $K$ and $H$ commute, the eigenfunction of $K$ is also eigenfunction of
the $H$. Therefore it is worth to obtain eigenfunction of $K$. When we solve
the following eigenvalue equation%
\begin{equation}
K\left\vert n_{1,}n_{2}\right\rangle =\left( k+\frac{1}{2}\right)
\left\vert n_{1,}n_{2}\right\rangle ,  \label{Keigenval}
\end{equation}
in the Bargman-Fock space we obtain the following expression:
\begin{equation}
\psi (x,y)=x^{k}\phi \left( xy\right) \left\vert \uparrow
\right\rangle +x^{k+1}\phi \left( xy\right) \left\vert \downarrow
\right\rangle . \label{psixy}
\end{equation}%
where $\left\vert \uparrow \right\rangle $ stands for up state and
$\left\vert \downarrow \right\rangle $ stands for down state. The
eigenfunction of the Hamiltonian can be obtained from the relation
\begin{equation}
\left\vert n_{1},n_{2}\right\rangle =\Gamma _{2}^{-1}\Gamma
_{1}^{-1}\Lambda ^{-1}\psi (x,y).  \label{n1n2}
\end{equation}%
Substitution of (\ref{psixy}) into the Hamiltonian (\ref{newham})
by using the realization of bosons in (\ref{abgoesto}) leads to
the following set of one variable coupled differential equations
\begin{eqnarray}
\hbar \omega \left[ 2z\frac{d}{dz}+k+1+\frac{k\omega _{c}}{2\omega
}-\frac{\mu gB}{2\hbar \omega }-\frac{E}{\hbar \omega }\right]
\phi _{1}\left(
z\right) &&  \nonumber \\
+ \lambda _{R}\left[ k+1-\frac{m^{\ast }\omega z}{\hbar
}+z\frac{d}{dz}\right] \phi _{2}\left( z\right) &=&0
\label{equa1}
\end{eqnarray}
and
\begin{eqnarray}
\hbar \omega \left[ 2z\frac{d}{dz}+k+2+\frac{(k+1)\omega
_{c}}{2\omega }+ \frac{\mu gB}{2\hbar \omega }-\frac{E}{\hbar
\omega }\right] \phi _{2}\left(
z\right) &&  \nonumber \\
+ \lambda _{R}\left[ \frac{m^{\ast }\omega }{\hbar
}-\frac{d}{dz}\right] \phi _{1}\left( z\right) &=&0  \label{equa2}
\end{eqnarray}
where $z=xy$ and $E$ is the eigenvalues of the Hamiltonian $H$ and $\phi
_{1}\left( z\right) $ and $\phi _{2}\left( z\right) $ correspond up and down
eigenstates of the Hamiltonian $H$, respectively. Following the analysis of
\cite{tutunculer} which was constructed to obtain the solution of
the Rashba Hamiltonian, one can obtain the quasi-exact solution of
the differential equations. Here we present solution of the problem
in the framework of the AIM.

\subsection{Solution of the Rashba Hamiltonian}

In the previous section we have formulated the Rashba Hamiltonian
based on the two boson operators and we have discussed its
transformation to the one variable differential equation. In this
section we present the AIM which leads to solution of the
Hamiltonian (\ref{newham}). For the sake of simplicity let us take
$\hbar =m^{\ast}=\omega _{0}=1$, then the last expression given
above can easily be written as the two coupled differential
equation
\begin{eqnarray}
\phi _{1}^{\prime } &=&a_{0}\phi _{1}+b_{0}\phi _{2}  \nonumber \\
\phi _{2}^{\prime } &=&c_{0}\phi _{2}+d_{0}\phi _{1}
\label{difpsis}
\end{eqnarray}
where
\begin{eqnarray}
a_{0} &=&\frac{\omega \left( 2E+\lambda _{R}^{2}+Bg\mu -\omega _{c}k-2\omega
(k+1)\right) }{\left( 4\omega ^{2}z+\lambda _{R}^{2}\right) }  \nonumber \\
b_{0} &=&\frac{\lambda _{R}\left( -2E+Bg\mu +\omega _{c}(k+1)-2\omega
k+4\omega ^{2}z\right) }{2\left( 4\omega ^{2}z+\lambda _{R}^{2}\right) }
\nonumber \\
c_{0} &=&\frac{\omega z\left( 2E-Bg\mu -\omega _{c}(k+1\right) -2\omega
(k+2))-\lambda _{R}^{2}(1+k-\omega z)}{z\left( 4\omega ^{2}z+\lambda
_{R}^{2}\right) }  \label{a0tod0} \\
d_{0} &=&\frac{\lambda _{R}\left( 2E+Bg\mu -\omega _{c}k-2\omega
(k+1)-4\omega ^{2}z\right) }{2z\left( 4\omega ^{2}z+\lambda _{R}^{2}\right) }
\nonumber
\end{eqnarray}
As in the previous section, one can compute $a_{n},b_{n},c_{n}$
and $d_{n}$ using the relations given in (\ref{constr1}) by a
simple MATHEMATICA program. When $z=0$, the eigenvalues are
produced by the roots of the quantization condition. The computed
results of the iteration are reported in Table \ref{tab:b}.

As it is seen that the energy eigenvalue equation is easily
obtained by using AIM. This is the advantage of the AIM that gives
the eigenvalues directly by transforming the quantum optical
Hamiltonians in the form of the $2\times 2$ matrix differential
equation form. In a similar manner as in the previous section one
can also obtain eigenfunction of the Hamiltonian (\ref{equa1}) \
and (\ref{equa2}) by using the relation (\ref{wavefunctions}).


\section{Conclusion}

We study the AIM to solve the problem of an electron in a quantum dot in the
presence of both magnetic field and spin-orbit coupling. Our formulation
gives an accurate result for the eigenvalues of Rabi and Rashba
Hamiltonians. The suggested approach can easily be modified to solve some
other quantum optical problems.

Furthermore we have presented a transformation procedure that offers several
advantageous, especially if one wishes to describe the eigenvalues of the
bosonic Hamiltonians by using AIM. It is obvious that the technique
presented here have been used in a variety of problems to compute their
spectrums. We have presented the steps towards an extension of the AIM.

The technique given in this article can be extended in several ways. The
Hamiltonian of a quantum dot including position dependent effective mass may
be formulated and solved within the procedure given here. We hope that our
method leads to interesting results on the spin-orbit effects in quantum
dots in future research. Success of our analysis leads to the solution of
the wide range of physical systems.

One of the authors (O. \"{O}zer) is grateful to the Abdus Salam
International Centre for Theoretical Physics, Trieste, Italy, for
its hospitality.

\newpage\


\begin{table}[tp]
\caption{Eigenvalues of the Rabi Hamiltonian for different values of energy
state $n$, splitting constant $\protect\omega_{0}$, and coupling constant
$\protect\kappa$.}
\label{tab:a}
\vspace{5mm}
$%
\begin{tabular}{|l|l|l|l|l|l|l|}
\hline
$n$ & $\omega _{0}$ & $\kappa $=0.0 & $\kappa $=0.25 & $\kappa $=0.50 & $%
\kappa $=0.75 & $\kappa $=1.0 \\ \hline
& $0.0$ & 0.5 & 0.46875 & 0.375 & 0.21875 & 0.0 \\ \cline{3-7}
&  & 0.5 & 0.46875 & 0.375 & 0.21875 & 0.0 \\ \cline{2-7}\cline{3-7}
$0$ & $0.5$ & 0.0 & -0.015748 & -0.064513 & -0.151195 & -0.284922 \\
\cline{3-7}
&  & 1.0 & 0.8080812 & 0.5865567 & 0.3344402 & 0.0488330 \\
\cline{2-7}\cline{3-7}
& $1.0$ & -0.5 & -0.510489 & -0.542870 & -0.600162 & -0.688478 \\ \cline{3-7}
&  & 0.5 & 0.4494499 & 0.3117158 & 0.1101679 & -0.141057 \\ \hline
& $0.0$ & 1.5 & 1.46875 & 1.375 & 1.21875 & 1.0 \\ \cline{3-7}
&  & 1.5 & 1.46875 & 1.375 & 1.21875 & 1.0 \\ \cline{2-7}
$1$ & $0.5$ & 2.0 & 1.7359493 & 1.4482320 & 1.1416351 & 0.8206445 \\
\cline{3-7}
&  & 1.0 & 1.1601606 & 1.2791925 & 1.3249343 & 0.0488339 \\ \cline{2-7}
& $1.0$ & 1.5 & 1.4124109 & 1.1964923 & 0.9129976 & 0.5913838 \\ \cline{3-7}
&  & 1.5 & 1.5294031 & 1.5980360 & 1.6078394 & 1.2925094 \\ \hline
& $0.0$ & 2.5 & 2.46875 & 2.375 & 2.21875 & 2.0 \\ \cline{3-7}
&  & 2.5 & 2.46875 & 2.375 & 2.21875 & 2.0 \\ \cline{2-7}
$2$ & $0.5$ & 2.0 & 2.2317488 & 2.4052487 & 2.4171516 & 2.1405752 \\
\cline{3-7}
&  & 3.0 & 2.6813089 & 2.3498342 & 2.0376015 & 1.8229775 \\ \cline{2-7}
& $1.0$ & 2.5 & 2.5660788 & 2.6980532 & 2.5296180 & 2.0954608 \\ \cline{3-7}
&  & 2.5 & 2.3778828 & 2.1020231 & 1.8333934 & 1.8464607 \\ \hline
& $0.0$ & 3.5 & 3.46875 & 3.375 & 3.21875 & 3.0 \\ \cline{3-7}
&  & 3.5 & 3.46875 & 3.375 & 3.21875 & 3.0 \\ \cline{2-7}
$3$ & $0.5$ & 4.0 & 3.6359015 & 3.2763600 & 3.0202409 & 2.9667546 \\
\cline{3-7}
&  & 3.0 & 3.2858045 & 3.4872316 & 3.4040837 & 3.0257950 \\ \cline{2-7}
& $1.0$ & 3.5 & 3.3455073 & 3.0291733 & 2.9681181 & 3.1101317 \\ \cline{3-7}
&  & 3.5 & 3.6001795 & 3.7530103 & 3.4089357 & 2.9237568 \\ \hline
& $0.0$ & 4.5 & 4.46875 & 4.375 & 4.21875 & 4.0 \\ \cline{3-7}
&  & 4.5 & 4.46875 & 4.375 & 4.21875 & 4.0 \\ \cline{2-7}
$4$ & $0.5$ & 4.0 & 4.3305823 & 4.5395425 & 4.347928 & 3.9264039 \\
\cline{3-7}
&  & 5.0 & 4.5965321 & 4.2235326 & 4.070904 & 4.0877781 \\ \cline{2-7}
& $1.0$ & 4.5 & 4.6320470 & 4.7466147 & 4.294056 & 3.7849395 \\ \cline{3-7}
&  & 4.5 & 4.3150285 & 3.9930696 & 4.121917 & 4.2669272 \\ \hline
& $0.0$ & 5.5 & 5.46875 & 5.375 & 5.21875 & 5.0 \\ \cline{3-7}
&  & 5.5 & 5.46875 & 5.375 & 5.21875 & 5.0 \\ \cline{2-7}
$5$ & $0.5$ & 6.0 & 5.5615693 & 5.1903384 & 5.1469657 & 5.1455854 \\
\cline{3-7}
&  & 5.0 & 5.3692738 & 5.5672925 & 5.2808323 & 4.8635200 \\ \cline{2-7}
& $1.0$ & 5.5 & 5.2862655 & 5.0134462 & 5.2558821 & 5.2799593 \\ \cline{3-7}
&  & 5.5 & 5.6619147 & 5.7022888 & 5.1889935 & 4.7100710 \\ \hline
\end{tabular}%
$%
\end{table}

\newpage\

\begin{table}[t]
\caption{Eigenvalues of the Rashba Hamiltonian for different values of
external magnetic field magnitude, $B$, energy state $n$, and coupling
constant $\protect\kappa$.}
\label{tab:b}
\vspace{5mm}
$%
\begin{tabular}{|l|l|l|l|l|l|l|l|}
\hline
$k$ & $B$ & $n$ & $\kappa =0.0$ & $\kappa =0.25$ & $\kappa =0.5$ & $\kappa
=0.75$ & $\kappa =1.0$ \\ \hline
&  & $0$ & 0.0 & 0.0 & 0.0 & 0.0 & 0.0 \\ \cline{3-8}
&  & $1$ & 1.0 & 0.9393054 & 0.7737872 & 0.5320678 & 0.2330309 \\ \cline{3-8}
& 0 & $2$ & 2.0 & 1.9962478 & 1.9465668 & 1.7810714 & 1.4877948 \\
\cline{3-8}
&  & $3$ & 3.0 & 2.9429223 & 2.8214396 & 2.7072634 & 2.5694541 \\ \cline{3-8}
&  & $4$ & 4.0 & 3.9928041 & 3.9101840 & 3.6989847 & 3.4345727 \\ \cline{3-8}
0 &  & $5$ & 5.0 & 4.9462371 & 4.8526593 & 4.7537468 & 4.5359330 \\
\cline{2-8}
&  & $0$ & 1.5 & 1.5 & 1.5 & 1.5 & 1.5 \\ \cline{3-8}
&  & $1$ & 0.5 & 0.4776216 & 0.4097986 & 0.2944479 & 0.1280664 \\ \cline{3-8}
& 3/2 & $2$ & 3.0 & 2.8877702 & 2.6235301 & 2.2882565 & 1.9132967 \\
\cline{3-8}
&  & $3$ & 4.0 & 4.0892712 & 4.2713119 & 4.2790928 & 3.8467145 \\ \cline{3-8}
&  & $4$ & 5.5 & 5.3107639 & 4.9316662 & 4.6839650 & 4.8228044 \\ \cline{3-8}
&  & $5$ & 6.5 & 6.6655604 & 6.9171192 & 6.5731725 & 6.0024178 \\ \hline
&  & $0$ & -1.0 & -1.0 & -1.0 & -1.0 & -1.0 \\ \cline{3-8}
&  & $1$ & 2.0 & 1.8847720 & 1.6082091 & 1.2505849 & 0.8439143 \\ \cline{3-8}
& 0 & $2$ & 3.0 & 3.0463969 & 3.0303266 & 2.7527743 & 2.3150943 \\
\cline{3-8}
&  & $3$ & 4.0 & 3.8955162 & 3.7596438 & 3.7797919 & 3.6831314 \\ \cline{3-8}
&  & $4$ & 5.0 & 5.0363268 & 4.9398058 & 4.6186235 & 4.3781510 \\ \cline{3-8}
&  & $5$ & 6.0 & 5.9052818 & 5.8399949 & 5.8328977 & 5.5201455 \\ \cline{2-8}
1 &  & $0$ & 1.0 & 1.0 & 1.0 & 1.0 & 1.0 \\ \cline{3-8}
&  & $1$ & 2.5 & 2.4555260 & 2.3240428 & 2.1107652 & 1.8227033 \\ \cline{3-8}
& 3/2 & $2$ & 5.0 & 4.8690809 & 4.5668665 & 4.1837412 & 3.7484901 \\
\cline{3-8}
&  & $3$ & 6.0 & 6.1078281 & 6.3230999 & 6.2570599 & 5.7506451 \\ \cline{3-8}
&  & $4$ & 7.5 & 7.2943805 & 6.8910658 & 6.7283380 & 6.9481530 \\ \cline{3-8}
&  & $5$ & 8.5 & 8.6817766 & 8.9384866 & 8.5213644 & 7.9201055 \\ \hline
\end{tabular}%
$%
\end{table}


\begin{thebibliography}{99}

\bibitem{prinz} G. A. Prinz, Science {\bf 282}, 1660 (1998).

\bibitem{wolf} S. A. Wolf, D. D. Awschalom, R. A. Buhrman, J. M. Daughtan,
S. vov Moln\'{a}r, M. L. Roukes, A. Y. Chtchelkanova and D. M.
Treger, Science {\bf 294}, 1488 (2001).

\bibitem{winkler} R. Winkler, Phys. Rev. B {\bf 62}, 4254 (2000).

\bibitem{wang} X. F. Wang, P. Vasilopoulos, F. M. Peeters, Appl. Phys. Lett. {\bf 80}, 1400 (2002).

\bibitem{band} S. Bandyopadhyay, Phys. Rev. B {\bf 61}, 13813 (2000).

\bibitem{rashba} Y. A. Bychkov and E. I. Rashba, J. Phys. C {\bf 17}, 6039 (1984).

\bibitem{das} S. Datta and B. Das, Appl. Phys. Lett. {\bf 56}, 665 (1990).

\bibitem{tutunculer} H. T\"{u}t\"{u}nc\"{u}ler, R. Ko\c{c} and E. Ol\u{g}ar, J. Phys. A {\bf 37}, 11431-11438 (2004).

\bibitem{judd} B. R. Judd, J. Phys. C {\bf 12}, 1685 (1979).

\bibitem{dresselhaus} G. Dresselhaus, Phys. Rev. {\bf 100}, 580 (1955).

\bibitem{jaynes} E. T. Jaynes and F. W. Cummings, Proc. IEEE {\bf 51}, 89 (1963).

\bibitem{rashbaei} E. I. Rashba, Sov. Phys. Solid State {\bf 2}, 1109 (1960).

\bibitem{tur} E. A. Tur, Opt. Spectrosc. {\bf 89}, 574 (2000).

\bibitem{tsit} E. Tsitsishvili, G. Lozano and A. O. Gogolin, Phys. Rev. B {\bf 70}, 115316 (2004).

\bibitem{frus} D. Frustaglia and K. Richter, Phys. Rev. B {\bf 69}, 235310 (2004).

\bibitem{koc1} R. Ko\c{c}, H. T\"{u}t\"{u}nc\"{u}ler, M. Koca and E. K\"{o}rc\"{u}k, Prog. Theor. Phys. {\bf 110}, 399 (2003);
 R.Ko\c{c}, M. Koca, H. T\"{u}t\"{u}nc\"{u}ler, J. Phys. A: Math. Gen. {\bf 35}, 9425-9430 (2002).

\bibitem{koc2} H. T\"{u}t\"{u}nc\"{u}ler and R. Ko\c{c}, PRAMANA J. Phys. {\bf 62}, 993-1005  (2004).

\bibitem{gover} M. Governale, Phys. Rev. Lett. {\bf 89}, 206802 (2002).

\bibitem{koc3} H. T\"{u}t\"{u}nc\"{u}ler and R. Ko\c{c}, Turk. J. Phys. {\bf 28}, 145-153  (2004);
R. Ko\c{c}, H. T\"{u}t\"{u}nc\"{u}ler, M. Koca and E. Ol\u{g}ar,
Ann. Phys. {\bf 319}, 333-347  (2005).

\bibitem{reik} H. G. Reik, M. E. St\"{u}lze and M. Doucha, J. Phys. A {\bf 20}, 6327 (1987).

\bibitem{loorits} V. Loorits, J. Phys. C {\bf 16}, L711 (1983).

\bibitem{ciftci1} H. \c{C}iftci, R. L. Hall and N. Saad, J. Phys. A {\bf 36}, 11807-11816 (2003).

\bibitem{sous} A. J. Sous, Chinese J Phys. {\bf 44}, 167 (2006).

\bibitem{ciftci2} H. \c{C}iftci, R. L. Hall, N. Saad, J. Phys. A {\bf 38}, 1147-1155 (2005).

\bibitem{ciftci3} H. \c{C}iftci, R. L. Hall and N. Saad, Phys. Rev. A {\bf 72}, 022101 (2005).

\bibitem{barakat1} T. Barakat, K. Abodayeh and A. Mukheimer, J. Phys. A {\bf 38} 1299-1304 (2005);
T. Barakat, Phys. Lett. A {\bf 344}, 411 (2005).

\bibitem{amore} F. M. Fern\`{a}ndez, J. Phys. A {\bf 37}, 6173--6180 (2004).

\bibitem{bayrak} O. Bayrak and I. Boztosun, J. Phys. A {\bf 39}, 6955-6964 (2006).
\end{thebibliography}
\end{document}